# Ocean Surface Roughness from Satellite Observations and Spectrum Modeling of Wind Waves

Paul A. Hwang[a]

[a] Remote Sensing Division, U. S. Naval Research Laboratory, Washington, DC (retired)

*Corresponding author:* Dr. Paul A. Hwang, hwangpaulbridge@gmail.com

ABSTRACT

Many wind wave spectrum models provide excellent wave height prediction given the input of wind speed and wave age. Their quantification of the surface roughness, on the other hand, varies considerably. The ocean surface roughness is generally represented by the mean square slope, its direct measurement in open ocean remains a challenging task. Microwave ocean remote sensing from space delivers ocean surface roughness information. Satellite platforms offer global coverage in a broad range of environmental conditions. This paper presents lowpass mean square slope (LPMSS) data obtained by spaceborne microwave altimeters and reflectometers operating at L, Ku, and Ka bands (about 1.6, 14, and 36 GHz). The LPMSS data represent the spectrally integrated ocean surface roughness with 11, 95, and 250 rad/m upper cutoff wave numbers, the maximum wind speeds are 80, 29, and 25 m/s, respectively. A better understanding of the ocean surface roughness is important to the goal of improving wind wave spectrum modeling. The analysis presented in this paper shows that over two orders of magnitude of the wave number range (0.3 to 30 rad/m), the spectral components follow a power function relating the dimensionless spectrum and the ratio between wind friction velocity and wave phase speed. The power function exponent is 0.38, which is considerable smaller than unity as expected from the classical equilibrium spectrum function. It may suggest that wave breaking is not only an energy sink but also a source of roughness generation covering a wideband of wavelengths about 20 m and shorter.

SIGNIFICANCE STATEMENT

This paper presents lowpass mean square slope (LPMSS) data obtained by spaceborne microwave altimeters and reflectometers operating at L, Ku, and Ka bands (about 1.6, 14, and 36 GHz). The LPMSS data represent the spectrally integrated ocean surface roughness with 11, 95, and 250 rad/m upper cutoff wave numbers, the maximum wind speeds are 80, 29, and 25 m/s, respectively. A better understanding of the ocean surface roughness is important to the goal of improving wind wave spectrum modeling that is critical to the investigation of air-sea interaction and ocean remote sensing. The analysis presented in this paper suggests that wave breaking is not only an energy sink but also a source of generating surface roughness covering a wideband of wavelengths about 20 m and shorter.

## 1. Introduction

The ocean surface roughness is of great interest to the investigation of air-sea interaction and ocean remote sensing. It is generally quantified by the ocean surface mean square slope (MSS) $s^2$, which is mainly contributed by short surface waves because of the $k^2$ weighting, where $k$ is wave number. Making measurements of MSS in the open ocean remains a serious technical challenge even today, and the airborne optical measurements in clean and slicked waters conducted more than 70 years ago in 1951 (Cox and Munk 1954a,b) remain the most celebrated MSS data for many decades, they are referred to as the C54 data sets in this paper. The maximum wind speeds are about 14 and 10 m/s in the clean and slicked data sets, respectively. (Unless otherwise specified, the data source or data set is identified by the initial of the lead author last name appended with the last two digits of publication year throughout this paper.)

Natural slicks usually cannot maintain a continuous coverage over a large area in winds above approximately 3 m/s. The C54 slicked-water measurements in winds up to 10 m/s is achieved by discharging a mixture containing crank case oil, diesel oil, and fish oil to produce a contiguous artificial slick field, about 200 gallons for a square area 2000 ft (610 m) a side. The application of artificial slick suppresses short waves, which are estimated to be less than about 0.3 m in wavelength, and the measured surface roughness is the lowpass MSS (LPMSS) $s_f^2$ with an upper cutoff wave number $k_u$ about 20 rad/m. The modification of surface boundary conditions by the artificial slick and the associated effects on wind wave generation are discussed in C54.

Microwave remote sensing is an environment-friendly way to obtain LPMSS and the non-contact operation does not disturb the air-sea interface. Electromagnetic (EM) theories of microwave scattering from rough surfaces and ocean scattering measurements have narrowed down the $k_u$ range to be between about $k_r/6$ and $k_r/3$ (Brown 1978; Plant 1986; Jackson et al. 1992; Voronovich and Zavorotny 2001; McDaniel 2001; Freilich and Vanhoff 2003; Warnick et al. 2005; Johnson et al. 2007), where $k_r$ is the EM (radar) wave number. Using different EM frequencies, LPMSS with different $k_u$ values can be obtained. Several airborne measurements at C (Hauser et al. 2008), Ku (Jackson et al. 1992), and Ka (Walsh et al. 1998; Vandemark et al. 2004) bands have been reported and the wind speed coverage extends to about 20 m/s. Over the years, microwave altimeters and reflectometers operating at different frequencies have been employed in satellite remote sensing of ocean surface winds. A recent review of the L- (Katzberg and Dunion 2009; Katzberg et al. 2013; Gleason 2013; Gleason et al. 2018; Balasubramaniam and Ruf 2020), Ku- (Hwang et al. 1998, 2002; Freilich and Vanhoff 2003; Ribal and Young 2019), and Ka-band (Lillibridge et al. 2014; Guerin et al. 2017) measurements is given in Hwang et al. (2021). Here the LPMSS results derived from those satellite sensors are presented. With the representative EM frequencies of 1.575, 13.6, and 35.75 GHz in the satellite data, the $k_u$ values are about 11, 95 and 250 rad/m, respectively (calculated by $k_r/3$, and to be further discussed in Section 2a). The maximum wind speeds in the L, Ku, and Ka data sets are respectively 80, 29, and 25 m/s, which represent a considerable expansion of the wind speed coverage, especially for the L band.

Section 2 presents a brief review of EM reflectometry and LPMSS. Using the forward computation relating the altimeter and reflectometer normalized radar cross section (NRCS) $\sigma_0$ and surface wind speed $U_{10}$, lookup tables (LUTs) are generated for retrieving LPMSS from observed NRCS. Section 3 discusses the related subjects of long- and short-wave spectrum functions, with emphasis on the ocean surface roughness and especially on the L band LPMSS, the contributions of surface roughness analyses toward the goal of improving wind wave spectrum modeling, and implications on wave dynamics. Section 4 is a summary.

## 2. LPMSS and Reflectometry

### *a. Specular point theory*

According to the specular point theory (SPT), the $\sigma_0$ is proportional to the product of the average number of specular points $n_A$ illuminated by the EM waves and the geometric mean of the two principal radii of curvature $\langle |r_1 r_2| \rangle$ of those specular points (Kodis 1966; Barrick 1968a,b). Closed-form solutions of $n_A$ and $\langle |r_1 r_2| \rangle$ are derived (Barrick 1968a,b),

$$n_A = \frac{7.255}{\pi^2 l^2} exp\left(\frac{-tan^2 \gamma}{s_f^2}\right) \quad (1)$$

$$\langle |r_1 r_2| \rangle = \frac{0.138 \pi l^2}{s_f^2} sec^4 \gamma \quad (2)$$

where $l$ is the correlation length between surface points $\zeta(x, y)$ and $\zeta(x', y')$ separated by a horizontal distance $\delta = [(x - x')^2 + (y - y')^2]^{1/2}$ assuming a Gaussian distribution of the correlation coefficient as $\delta \to 0$, and $\gamma$ is the surface tilting angle at the specular point. The numerical constants in (1) and (2) result from carrying out triple integrals of ($\zeta_{xx}$, $\zeta_{xy}$, $\zeta_{yy}$) functions. The NRCS scattered from a rough surface is then given by:

$$\sigma_{0pq}(\theta_s, \theta_i, \gamma) = |R_{pq}|^2 \frac{sec^4 \gamma}{s_f^2} exp\left(\frac{-tan^2 \gamma}{s_f^2}\right) \quad (3)$$

where subscripts $p$ and $q$ are the incident and scattered polarizations, respectively, $\theta_i$ and $\theta_s$ are the incidence and scattering angles, $R_{pq}$ is the Fresnel reflection coefficient, $s_f^2$ is the LPMSS (Valenzuela 1978), and $\tan\gamma$ is the surface slope at the specular point, which is a function of incidence and scattering angles

$$\tan\gamma = \frac{(\sin^2\theta_i - 2\sin\theta_i \sin\theta_s \cos\phi_s + \sin^2\theta_s)^{1/2}}{\cos\theta_i + \cos\theta_s} \quad (4)$$

where $\phi_s$ is the azimuth angle of the scattering vector; the solution is formulated with $\phi_i = 0$ and $\phi_i$ is the azimuth angle of the incidence vector. The solution form in (3) indicates an exponential attenuation with respect to the facet (specular point) tilting angle. This is an important consideration for nadir (altimeter) backscattering return ($\theta_i = \theta_s = 0$, $\phi_s = \pi$) and forward specular reflection ($\theta_i = \theta_s$, $\phi_s = 0$), for which $\gamma = 0°$. Because of the nature of nonuniform distribution of ocean surface roughness and sporadic or intermittent generation of short waves, the received scattering is the sum of contributions from all roughness patches in the antenna footprint, and the tilting of various scattering roughness patches needs to be considered. The effective patch size is scaled by the EM wavelength, for example, for a 2-cm Ku band illumination, a square patch a few meters long on each side would contain many reflecting facets. The patch can ride at any location on the ocean waves that tilt at different angles, thus the local incidence, scattering, and specular-point angles are modified. Consequently, patches with statistically identical roughness $s_f^2$ at different phases of long waves have different contributions to the received overall scattering return, and $s_f^2$ of a roughness patch alone is not sufficient to characterize its contribution to the overall scattering. In essence, surface waves several times longer than the EM wavelength are serving two roles: they are the surface roughness elements and they are also the agent that facilitates the modification of local incidence angle; further discussion on this issue is given in Section 3.d.

Applying the tilted Bragg (two-scale) scattering concept to account for the local incidence angle modification (Valenzuela 1978), the NRCS for tilted specular reflection ($\theta_s = \theta_i$) is then (Hwang et al. 2021)

$$\sigma_{0pq}(\theta_s = \theta_i) = \int_{-\infty}^{\infty}\int_{-\infty}^{\infty} |R_{pq}|^2 \frac{\sec^4\gamma}{s_f^2} \exp\left(\frac{-\tan^2\gamma}{s_f^2}\right) p\left(\tan\gamma_x, \tan\gamma_y\right) d\tan\gamma_x d\tan\gamma_y \quad (5)$$

where $p\left(\tan\gamma_x, \tan\gamma_y\right)$ is the probability distribution function (pdf) of the tilting surface slope components. To carry out the computation, a surface wave spectrum is needed to generate the input $s_f^2$. The wave spectrum functions (G18 and H18) used in Hwang et al. (2021) will be further discussed in Section 3. The pdf is assumed to be Gaussian. The Fresnel reflection coefficient is determined by the EM frequency and the incidence and scattering angles, with modification accounting for the air entrainment from breaking-produced whitecaps (Hwang 2012; Hwang et al. 2019a,b). In Hwang et al. (2021), computations with $k_u = k_r/3$ and $k_r/5$ are presented. The results with $k_u = k_r/3$, as summarized in Figure 1, are clearly in better agreement with satellite measurements and they are used in this paper.

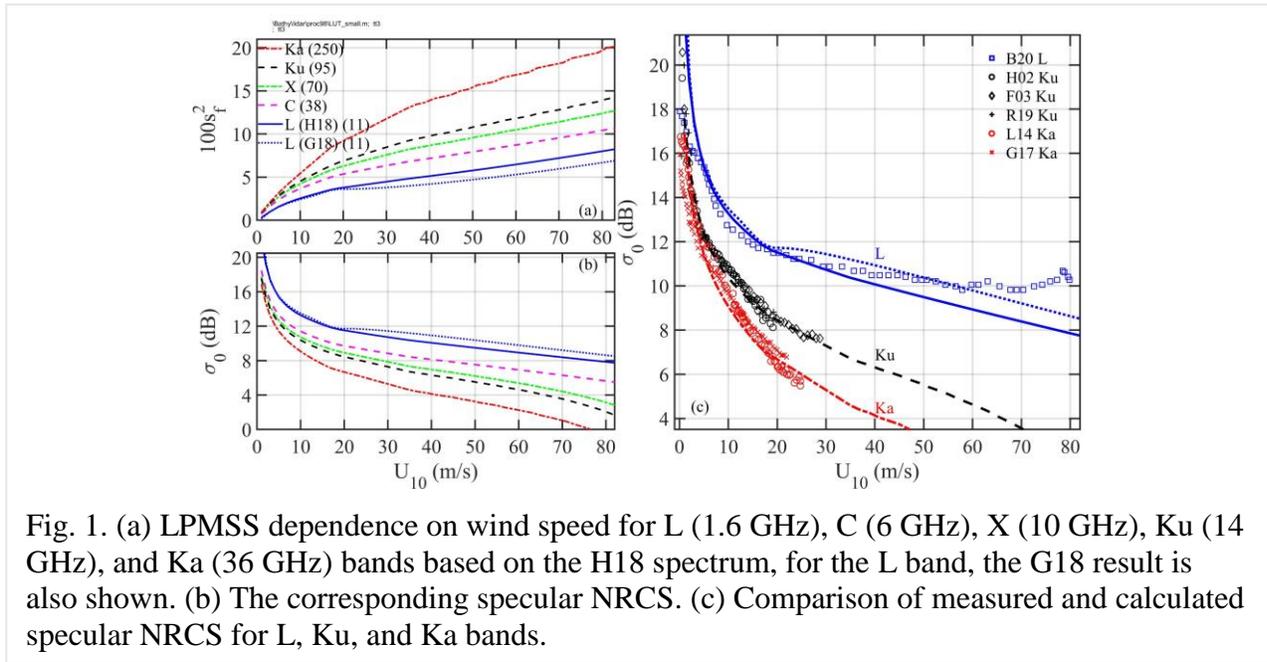

Fig. 1. (a) LPMSS dependence on wind speed for L (1.6 GHz), C (6 GHz), X (10 GHz), Ku (14 GHz), and Ka (36 GHz) bands based on the H18 spectrum, for the L band, the G18 result is also shown. (b) The corresponding specular NRCS. (c) Comparison of measured and calculated specular NRCS for L, Ku, and Ka bands.

Figure 1a shows the computed $s_f^2$ for several EM frequencies (Ka: 35.75 GHz, Ku: 13.6 GHz, X: 10 GHz, C: 5.5 GHz, and L: 1.575 GHz), the corresponding $k_u$ of each frequency is given in parentheses in the legend. For the L band, results from both G18 and H18 spectra are displayed. For C band and higher frequencies, the results are based on the H18 spectrum function. Figure 1b shows the corresponding specular NRCS computed with (4). There are now many satellite NRCS data sets of Ku and Ka band altimeters and L band reflectometers reported in the literature as mentioned in the Introduction. Figure 1c shows the good agreement (within 1 dB for $U_{10}$ up to about 60 m/s) between EM computations and the following satellite data sets: L (Balasubramaniam and Ruf 2020), Ku (Hwang et al. 2002; Freilich and Vanhoff 2003; Ribal and Young 2019), and Ka (Lillibridge et al. 2014; Guerin et al. 2017).

### b. LPMSS retrieval

The good agreement between EM solutions and satellite data is encouraging and we can use the EM solutions as the geophysical model function (GMF) $s_f^2(\sigma_0)$ for retrieving LPMSS from NRCS (Figure 2a). The results can also be given as a lookup table (LUT) of $(U_{10}, \sigma_0, s_f^2)$ for each microwave frequency. The LUTs of L, C, X, Ku, and Ka frequencies, with wind speeds between 1 and 99 m/s in steps of 1 m/s, are given in the supplemental material (LUT_U10_NRCS_MSS.pdf). The wind speed or LPMSS can then be obtained from NRCS by linear interpolation using the LUTs. Figure 2b shows the retrieved Ka-, Ku-, and L-band LPMSSs plotted against the wind speed of each satellite data set. For Ku and Ka bands, the retrieved results from different satellite data sets differ only slightly, within about 0.01 $s_f^2$ magnitude. For the L band, retrieved values using both G18 and H18 $s_f^2(\sigma_0)$ curves are presented. The two sets are similar for $U_{10}$ less than about 20 m/s. At higher wind speeds, the difference is mostly within about 0.01 $s_f^2$ magnitude as well. The optical C54 data sets in clean and slicked waters are displayed with magenta markers in Figure 2b for comparison. The clean-water data are the approximate upper bound of the microwave data, and the slicked-water data are close to the microwave L band results. Also shown in the figure is an optical data set (B06) from spaceborne measurements (Bréon and Henriot 2006), the spaceborne optical data are essentially the same as the C54 clean water results.

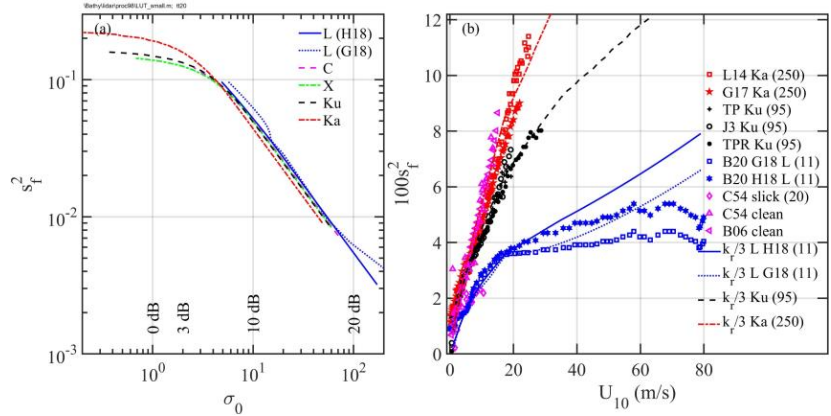

Fig. 2. (a) EM solutions of $s_f^2(\sigma_0)$ serving as the GMF for retrieving LPMSS from specular NRCS. (b) The L, Ku, and Ka bands LPMSS retrieved from their NRCS and plotted as functions of wind speed. Also plotted are the C54 and B06 optical data (magenta markers). Smooth lines show the spectrum-computed LPMSS.

Here it is pointed out that the C54 clean water data may not represent the true optical limit of the ocean surface roughness because of the limitations in the assumptions of the slope statistics, factors other than the slope probability entering into the problem of determining the optical density of the photometric measurements of glitter-produced blobs on the film, and "Sunlight scattered from submerged particles and reflected skylight produce a background in which the sun glitter from large and infrequent slopes is lost" (Cox and Munk 1954a,b; Wentz 1976). Wentz (1976) concludes that the C54 clean water data provide a lower-bound estimate of the optical ocean surface roughness, and several papers (Wu 1971; Pierson and Stacy 1973; Wentz 1975) are cited to support his assessment.

## 3. Discussion

### a. Spectrum functions

The surface wave spectrum models used in the EM computation presented in Section 2 can be divided into two parts: the short waves and the long waves.

The short-wave portion, here referred to as the H15 spectrum function, is based on the power-function similarity relationship observed from analyzing field measurements of short-wave spectra

$$B(k) = A(k) \left(\frac{u_*}{c}\right)^{a(k)} \quad (6)$$

where $B(k) = k^3 S(k)$ is the dimensionless wave spectrum, $S(k)$ is the wave elevation spectrum, $u_*$ is the wind friction velocity, $c$ is the phase speed of the $k$ wave component, $A(k)$ and $a(k)$ are coefficients derived from empirical fitting of field data. The field measurements of short-wave spectra are collected by high-resolution wave wires mounted on a free-drift instrument platform to minimize Doppler frequency shift in the measured spectra (Hwang and Wang 2004a). The resolvable $k$ range is between about 0.6 and 326 rad/m, and the wind speed range is between 2 and 14 m/s (Hwang and Wang 2004b; Hwang 2005). Analytical asymptotic functions are developed to expand the wave number coverage from 0 to ∞ (Hwang 2008, 2011). The L-, C-, and Ku-band scatterometer GMFs from airborne and satellite measurements are used to refine the $A(k)$ and $a(k)$ functions, and to expand the wind speed coverage to tropical cyclone (TC) conditions (Hwang et al. 2013; Hwang and Fois 2015). The supplemental material (H15SpectrumFormulationNote.pdf) summarizes the H15 spectrum formulation, including a couple of post-2015 modifications prompted by replacing the low wave number portion with G18 as partially explained in Hwang (2019).

With $U_{10}$ typically serving as the wind input, a formula of drag coefficient $C_{10}$ is needed to compute $u_*$. Microwave brightness temperature measurements acquired in mild to TC conditions are used to evaluate various $C_{10}$ formulas, first with the large collection of Stepped Frequency Microwave Radiometer (SFMR) data (Klotz and Uhlhorn 2014), and subsequently extending to other sources including more refined and expanded SFMR (Sapp et al. 2019), WindSAT (Meissner and Wentz 2009), Aquarius, Soil Moisture Active Passive (SMAP), airborne Passive–Active L-band System (PALS) (Yueh et al. 2016; Meissner et al. 2017), and Soil Moisture and Ocean Salinity (SMOS) (Reul et al. 2016). The recommended $C_{10}$ formula as from the analyses is (Hwang 2018, 2019; Hwang et al. 2019a,b)

$$C_{10} = \begin{cases} 10^{-4}(-0.0160U_{10}^2 + 0.967U_{10} + 8.058), & U_{10} \leq 35 \ m/s \\ 2.23 \times 10^{-3}(U_{10}/35)^{-1}, & U_{10} > 35 \ m/s \end{cases} \quad (7)$$

The long-wave portion, here referred to as the G18 spectrum function, follows the traditional development of wind wave spectrum models and accepts a variable spectral slope (Pierson and Moskowitz 1964; Hasselmann et al. 1973, 1976; Donelan et al. 1985; Young 2006; Hwang et al. 2017; Hwang and Fan 2018)

$$S(\omega) = \left\{ \alpha g^2 \omega_p^{-5} \ \Omega^{-s_\omega} exp\left[-\frac{5}{4}\Omega^{-4}\right]\right\} \gamma^\Gamma \quad (8)$$

where the first curly-braced term on the right hand side is a modified P function (Pierson and Moskowitz 1964) with a variable spectral slope $s_\omega$, the second term with $\Gamma = exp[-(1-\Omega)^2/(2\sigma^2)]$ is the peak enhancement function (Hasselmann et al. 1973, 1976); $g$ is the gravitational acceleration, $\Omega = \omega/\omega_p$, $\omega$ is the angular frequency, and subscript $p$ indicates the spectral peak component; $\alpha$, $\gamma$, and $\sigma$ are spectral parameters varying with $s_\omega$ and $\omega_\# = \omega_p U_{10}/g$. A comprehensive description of the G18 spectrum with the complete set of formulas is given in Hwang (2022a). The G18 function is formulated to produce good agreement in mild to TC conditions not only with observed significant wave height (Table 2 in Hwang et al. 2017; Figure 1 in Hwang 2022a) but also with observed L band LPMSS (Figures 7 - 9 in Hwang and Fan 2018; Figure 6 in Hwang 2020a). An empirical function of average spectral slope $s_\omega(U_{10})$ is obtained by matching the measured LPMSSs with those computed with the wave spectrum function:

$$s_\omega(U_{10}) = \begin{cases} 4.7, & U_{10} \leq 18 \ m/s \\ 4.7\left(\frac{U_{10}}{18}\right)^{1/8}, & U_{10} > 18 \ m/s \end{cases} \quad (9)$$

Figure 3 shows examples of H15 and G18 spectra computed with $U_{10}$ and $\omega_\#$ listed inside the square brackets in the figure legend. Large differences between H15 and G18 in the high wave number region are obvious. The H15 is formulated to give good agreement with L-, C-, and Ku-band scatterometer GMFs and radiometer brightness temperature measurements from L to Ka bands in mild to TC wind conditions (Figures 7-9 in Hwang et al. 2013; Figure 5 in Hwang and Fois 2015; Figures 2-7 in Hwang et al. 2019a; Figure 3 in Hwang 2020b). The spaceborne scatterometers and radiometers operate at angles far away from specular reflection, typically between about 40° and 55° from nadir, the received backscattered and emitted signals from the sea surface are dominated by Bragg resonance from surface roughness components with wavelengths in the neighborhood of the EM wavelengths (Crombie 1955; Wright 1966, 1968; Johnson and Zhang 1999). For reference, the nominal EM wave numbers for L, C, Ku, and Ka bands: $k_{rL}$, $k_{rC}$, $k_{rKu}$, and $k_{rKa}$, respectively, are marked with blue vertical lines in Figure 3b.

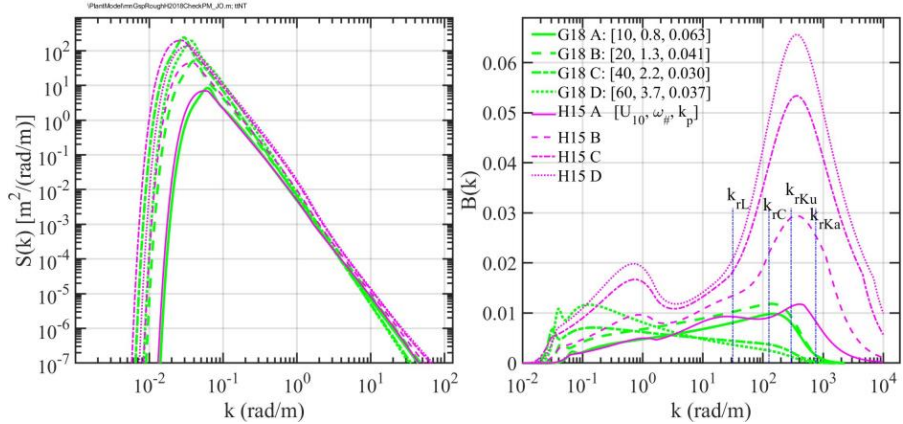

Fig. 3. Examples of H15 and G18 spectra: (a) $S(k)$, and (b) $B(k)$. Vertical lines show nominal L-, C-, Ku-, and Ka-band EM (radar) wave numbers $k_{rL}$, $k_{rC}$, $k_{rKu}$, and $k_{rKa}$, respectively.

For the long-wave portion, say, $k < \sim 10$ rad/m, H15 is close to G18 in low to moderate winds, e.g., compare the 10 m/s spectra (solid curves). The difference between H15 and G18 becomes progressively larger as the wind speed increases further, e.g., compare the 20 (dashed curves), 40 (dashed-dotted), and 60 (dotted curves) m/s spectra. For the EM computation presented in Section 2, a simple linear interpolation merges the H15 for short waves and G18 for long waves. The transition region is set to be between 1 and 4 rad/m (Hwang and Ainsworth 2020) based on the consideration that the apparent double-hump feature in the H15 spectrum function may be indicative of the dominance of wave breaking patch lengths at the valley between the two humps. The location of the valley, near $k = 2$ to 3 rad/m, suggests a strong signature in the couple-of-meters scale for the breaking wavelengths as observed in several field experiments of radar sea spikes (Liu et al. 1998; Frasier et al. 1998; Hwang et al. 2008). Detailed discussion on this issue is given in (Hwang 2007) and it will be revisited in Section 3c. The hybrid spectrum merging the H15 for short waves and G18 for long waves is referred to as the H18 spectrum. More detailed discussion on the H15 and H18 functions and their applications to active and passive microwave remote sensing are given in Hwang (2019).

## b. L band LPMSS and transition between long and short waves

Interests in exploring the global positioning system (GPS), or global navigation satellite systems (GNSSs) in general, for ocean remote sensing since the 1990s (Martín-Neira 1993; Katzberg and Garrison 1996) has led to an explosion of L-band LPMSS ($s_{fL}^2$) data (Katzberg and Dunion 2009; Katzberg et al. 2013; Gleason 2013; Gleason et al. 2018) with wind speed coverage extending to about 60 m/s (green and cyan markers in Figure 4). These data are derived from analyzing the Doppler properties of GPS reflectometry (GPSR). They have been used to determine the average spectral slope $s_\omega$ of the G18 spectrum function and an empirical $s_\omega(U_{10})$ relation (9) is obtained by matching the GPSR-derived and spectrum-computed $s_{fL}^2$ (Hwang and Fan 2018; Hwang 2020a). In return, it becomes feasible to compute $s_{fL}^2$ with the G18 spectrum function given only the $U_{10}$ and $\omega_\#$ input (Hwang 2021). This approach is used to investigate the $s_{fL}^2$ inside TCs with synthetic data generated by the G18 spectrum function coupled with the simulated $U_{10}$ and $\omega_\#$ computed from their similarity functions (Hwang and Ainsworth 2020). It is also used to quantify the azimuthal variation of $s_{fL}^2$ inside TCs by analyzing the G18 computations with the measured $U_{10}$ and $\omega_\#$ by hurricane hunters (Hwang 2022b). Dividing the TC coverage area into quarters ($-45° - 45°$, $45° - 135°$, $135° - 225°$, $225° - 315°$) and quadrants ($270° - 360°$, $0° - 90°$, $90° - 180°$, $180° - 270°$) with respect to the TC heading (angles increase counterclockwise), the average results in the four quarters and four quadrants are examined. Examples of the results in the front (F), left (L), back (B), and right (R) TC quarters are superimposed in Figure 4, the maximum wind speed in the hurricane hunter measurements is 67 m/s. For the same wind speed the $s_{fL}^2$ in the back quarter is about 12% to 20% higher than that in the front quarter in regions with $U_{10}$ greater than about 20 m/s; because of the influence from non-local waves the wind sea $\omega_\#$ values cannot be reliably determined for the hurricane hunter wave spectra in low winds ($U_{10}$ less than about 20 m/s, and all within 20 km from the TC center in the four hurricane hunter data sets examined). The $s_{fL}^2$ of left and right quarters are between those of the front and back quarters, and with much smaller differences. These spectrum-integrated $s_{fL}^2$ are in good agreement with the GPSR measurements. The analysis of spectrum-derived $s_{fL}^2$ indicates that the azimuthal variation can contribute significantly to the observed data scatter in the LPMSS data.

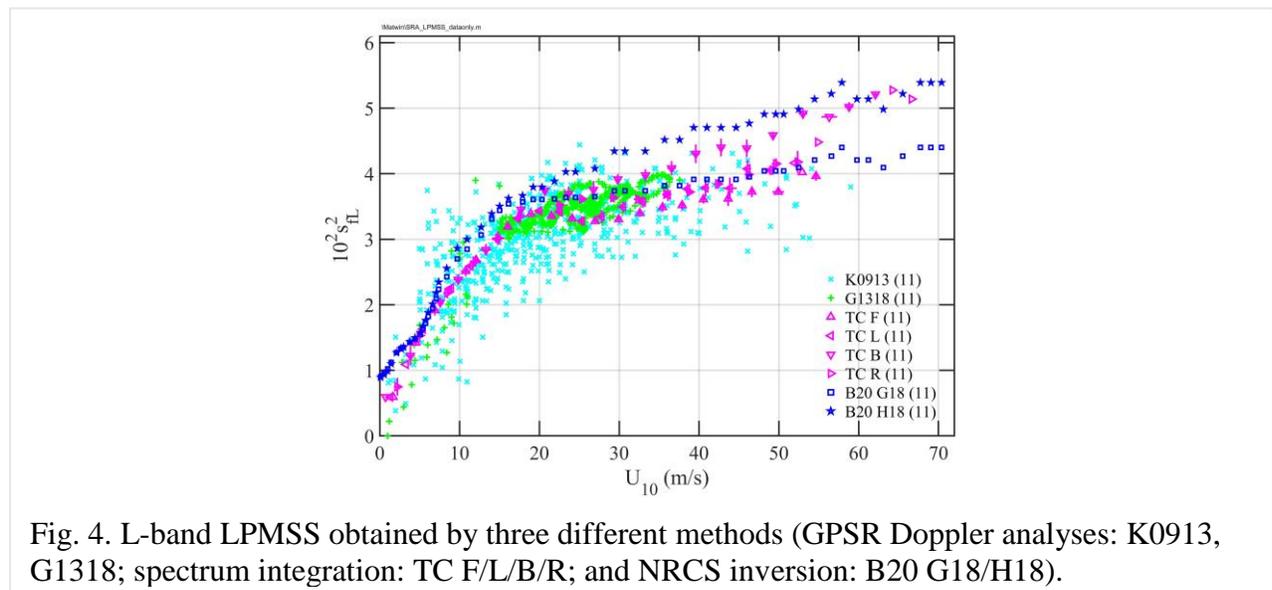

Fig. 4. L-band LPMSS obtained by three different methods (GPSR Doppler analyses: K0913, G1318; spectrum integration: TC F/L/B/R; and NRCS inversion: B20 G18/H18).

The $s_{fL}^2$ derived from the NRCS analysis presented in Section 2 are also added in Figure 4 (labeled B20); results shown with squares and stars are based on the G18 and H18 spectra respectively used to generate the input LPMSS in the EM solution (5). The $s_{fL}^2$ derived with the three different methods (GPSR Doppler analysis, wave spectrum integration, and NRCS inversion) are in good agreement. The NRCS-retrieved $s_{fL}^2$ with H18 spectrum function tend to be on the high side, and it suggests that the transition between H15 and G18 should move up from 1<$k$<4 rad/m to a range closer to 10 rad/m.

*c. Notable spectrum features and implications on wave dynamics*

With the current formulations, all wave components of the H15 spectrum increase monotonically with wind, which is consistent with scatterometer and radiometer observations. For the G18 spectrum, many wave components shorter than the spectral peak wavelength may decrease with wind speed greater than about 20 m/s (Figure 3b, green curves). The physical implication is explored.

Figure 5a show the computed G18 frequency spectra of wave elevation for a range of wind speeds between 5 and 70 m/s presented in the area-conserving semilogarithmic form, i.e., the area under each curve within a certain frequency range is the elevation variance integrated over the specified frequency range. Figure 5b shows the cumulatively integrated elevation variance $I(\omega) = \int_0^\omega S(\omega') \, d\omega'$ in normalized form. The wave development stage is represented by the inverse wave age $\omega_\# = \omega_p U_{10}/g = U_{10}/c_p$. Because $U_{10}$ and $\omega_\#$ are not independent variables, they should not be specified independently when modeling the wave spectra intending to represent realistic field observations over a wide wind speed range. The empirical relationship of the average windsea wave age used here is obtained from analyzing the wind and wave data combining a one-year-long recording in central Bering Sea and four hurricane hunter missions (Figure 5 in Hwang 2022a)

$$\omega_\# = 6.46 \times 10^{-4} U_{10}^2 + 8.77 \times 10^{-3} U_{10} + 0.822 \qquad (10)$$

Because of the sharp dropoff of wave energy away from the spectral peak, the wave elevation variance is contributed mainly by components near the spectral peak region: almost negligible from components with $\omega < 0.7\omega_p$, ~40% from $\omega < \omega_p$, and ~95% from $\omega < 2\omega_p$ (Figure 5b).

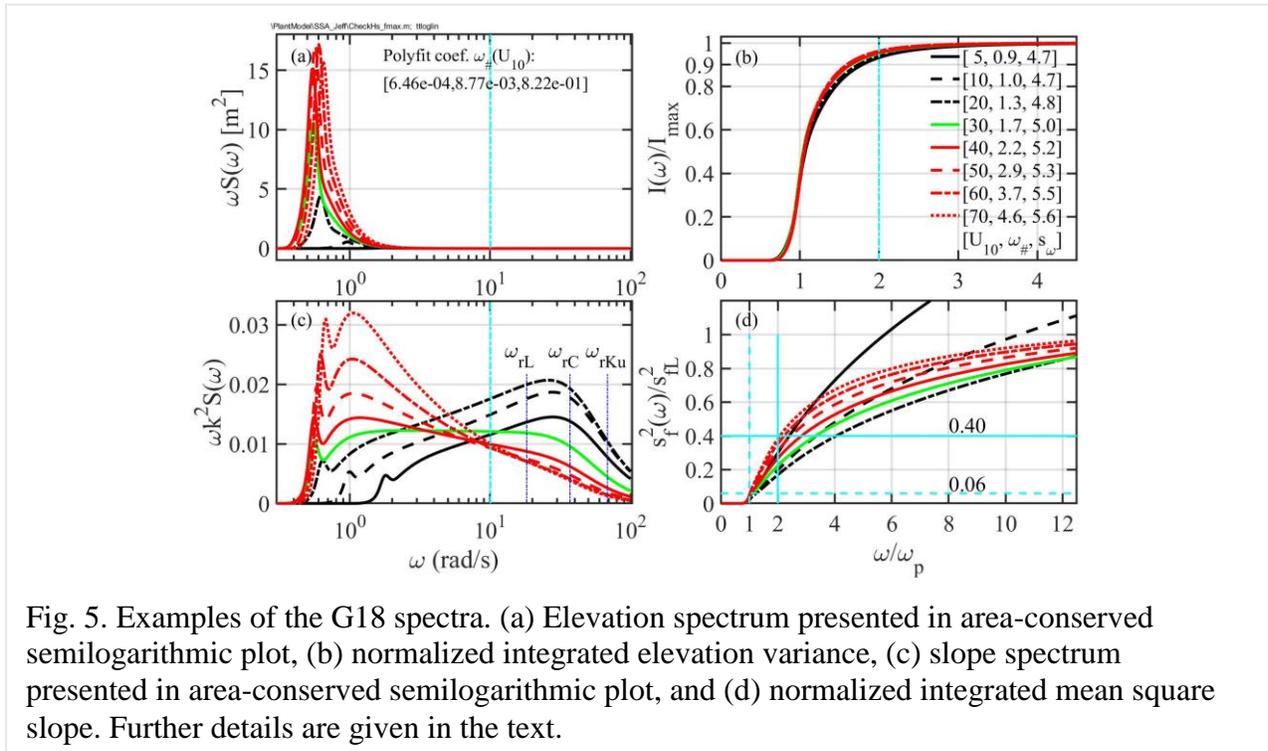

Fig. 5. Examples of the G18 spectra. (a) Elevation spectrum presented in area-conserved semilogarithmic plot, (b) normalized integrated elevation variance, (c) slope spectrum presented in area-conserved semilogarithmic plot, and (d) normalized integrated mean square slope. Further details are given in the text.

Figure 5 bottom row shows the slope spectra in the same formats as the elevation spectra, that is, Figure 5c gives the area-conserved semilogarithmic plot, and Figure 5d is the normalized cumulative LPMSS $s_f^2(\omega) = \int_0^\omega k^2 S(\omega') d\omega'$, here $s_{fL}^2 = \int_0^{11} k^2 S(k') dk'$ is used as the normalization factor. Although the spectrum computation is carried out to $\omega = 100$ rad/s, the applicable maximum frequency of the G18 spectrum function is estimated to be about $\omega_{max} = 10$ rad/s, which is about $k_{max} = 10$ rad/m, as marked by the vertical cyan lines in Figure 5a and Figure 5c. The components near the spectral peak are still important contributors to $s_{fL}^2$ but not as dominant as in the case of elevation variance, and the weighting clearly shifts to the higher frequency components: there is less than 6% contribution from components with $\omega < \omega_p$, and no more than 40% from $\omega < 2\omega_p$ (Figure 5d). The slope spectrum dependence on wind speed is complicated (Figure 5c). At $U_{10} = 30$ m/s (green curves), the average spectral slope $-s_\omega$ is $-5.0$ and the dimensionless slope spectrum has a flat region in the spectral tail (between about 2 and 20 rad/s, Figure 5c). Note: for gravity waves, $\omega k^2 S(\omega)$ as plotted in Figure 5c is $2B(k)$ because $S(k)\mathrm{d}k = S(\omega)\mathrm{d}\omega$ and $\omega^2 = gk$; $B(k) = k^3 S(k)$. For $U_{10} < 30$ m/s, there is a monotonic increase of all spectral components because $s_\omega$ is less than 5.0. For $U_{10} > 30$ m/s, the increasingly steepening of the spectral dropoff is a consequence of increasing $s_\omega$ beyond 5.0 as given by the empirical function (9). For convenience, $[U_{10}, \omega_\#, s_\omega]$ for each curve are given in the Figure 5b legend. The $s_\omega$ dependence on wind speed as defined by (9) yields a region in the neighborhood of $\omega = \sim 8$ rad/s where the spectral variation is least sensitive to the changing wind speed. The non-monotonic wind dependence in the spectral levels is not observed in microwave scatterometer and radiometer measurements that respond mainly to the resonance roughness elements with length scales close to the EM wavelengths. For reference, the L, C, and Ku EM wavelengths are converted to surface wave angular frequencies and shown with blue vertical lines in Figure 5c. Clearly, the G18 spectrum function should not be applied to $\omega$ much greater than about 10 rad/s where we have plenty of scatterometer and radiometer data showing the generally monotonic increasing trend with wind speed of the Bragg resonance wave spectral levels.

  In the hybrid H18 spectrum function used for the EM computation described in Section 2, the transition region merging the G18 and H15 is $1 < k < 4$ rad/m. The analyses presented in Sections 2b and 3b suggest that this transition region should be moved closer to $k = 10$ rad/m but below the L-band Bragg resonance wavelengths (using 20 rad/m as the conservative lower range, calculated with 1 GHz EM frequency and 30° incidence angle). Figure 6 shows the updated hybrid spectra with the transition region set at $8 < k < 16$ rad/m. The simple linear interpolation yields a smooth merging of the G18 and H15 functions for $U_{10}$ up to about 20 m/s (black and blue curves). At higher winds, the discontinuity between G18 and H15 in the transition region becomes more severe and the kinks at the two edges of the transition region are more pronounced (green and red curves). These computed spectra can serve as synthetic wind sea data with average wave development stage over a wide range of wind speeds for investigating the ocean wave properties, especially the ocean surface roughness.

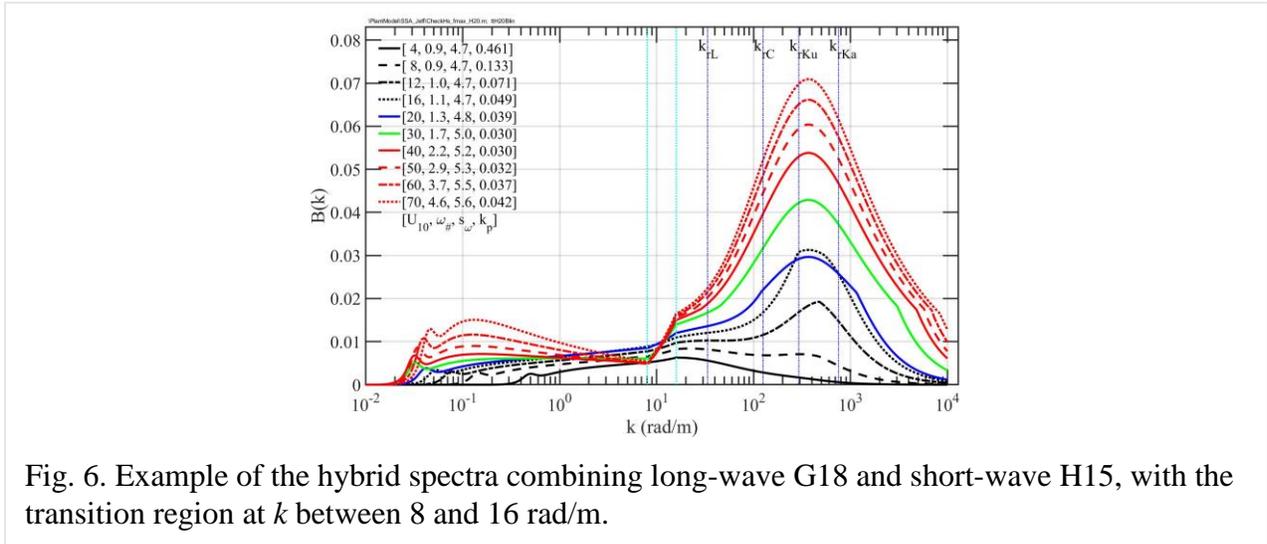

Fig. 6. Example of the hybrid spectra combining long-wave G18 and short-wave H15, with the transition region at *k* between 8 and 16 rad/m.

The design of the short-wave spectrum function (6) is prompted by the analyses of source function balance (Phillips 1984, 1985) that stress the importance of seeking out the *B*(*k*) functional representation, specifically its dependence on the wind forcing parameter $u_*/c$. Wave spectrum measurements focusing on centi- to deca-meter (cmDm) wavelengths collected in the ocean serve as the seed data of the $B(k) = A(k)(u_*/c)^{a(k)}$ similarity function of cmDm waves (Hwang and Wang 2004b; Hwang 2005, 2008, 2011). The same similarity relationship is also found in the analysis to retrieve Bragg resonance waves from L-, C-, and Ku-band scatterometer backscattering (Hwang et al. 2013; Hwang and Fois 2015). The connected blue, black, and red markers in Figure 7a show the similarity relationship applied to several wave spectral components with *k* ranging from 0.3 to 720 rad/m. The results represent a reformatting of the computed spectra shown in Figure 6. It is emphasized that Figures 6 and 7a present the identical set of computed wave spectra. The similarity function format: $B(u_*/c; k) = A(k)(u_*/c)^{a(k)}$ given in Figure 7a provides a more organized and trackable representation of cmDm waves. With the similarity function, the cmDm wave spectrum is determined by the *A*(*k*) and *a*(*k*) parameters; the characteristics of *A*(*k*) and *a*(*k*) may contain important information on the cmDm wave properties. For example, *A*(*k*) and *a*(*k*) analyses have been used to investigate the length scales of wave breaking patches, with the consideration that *B*(*k*), being computed with the surface waveform, represents the spectral decomposition of surface geometry; and wave breaking is a distinct characteristic of surface geometry (Hwang 2007).

The most noticeable feature of the spectra presented in Figure 7a is that: stretching over two orders of magnitude of the $k$ range (0.3 to 30 rad/m, blue and black markers), the results group tightly forming a simple power function

$$B(k) = 0.011(u_*/c)^{0.38}, \tag{11}$$

which is shown with the green solid line. Deviation from this simple function occurs for $U_{10}$ greater than about 20 m/s as seen in the 3 to 5 markers in each connected blue or black curve with higher $u_*/c$ values; the spectra are computed for $U_{10}$ = 4, 8, 12, 16, 20, 30, 40, 50, 60, and 70 m/s, for each curve, increasing $u_*/c$ represents increasing $U_{10}$. The short-wave spectrum function H15 is formulated with the $B(k) = A(k)(u_*/c)^{a(k)}$ similarity relation, so the outcome of spectral components $k$ from about 15 to 30 rad/m converging to (11) is not necessarily surprising. On the other hand, spectral components $k < 8$ rad/m are formulated in a quite different fashion and they are determined by $U_{10}$ and $\omega_\#$ (the G18 spectrum), so the converging to (11) of components $k$ from about 0.3 to 10 rad/m is rather intriguing.

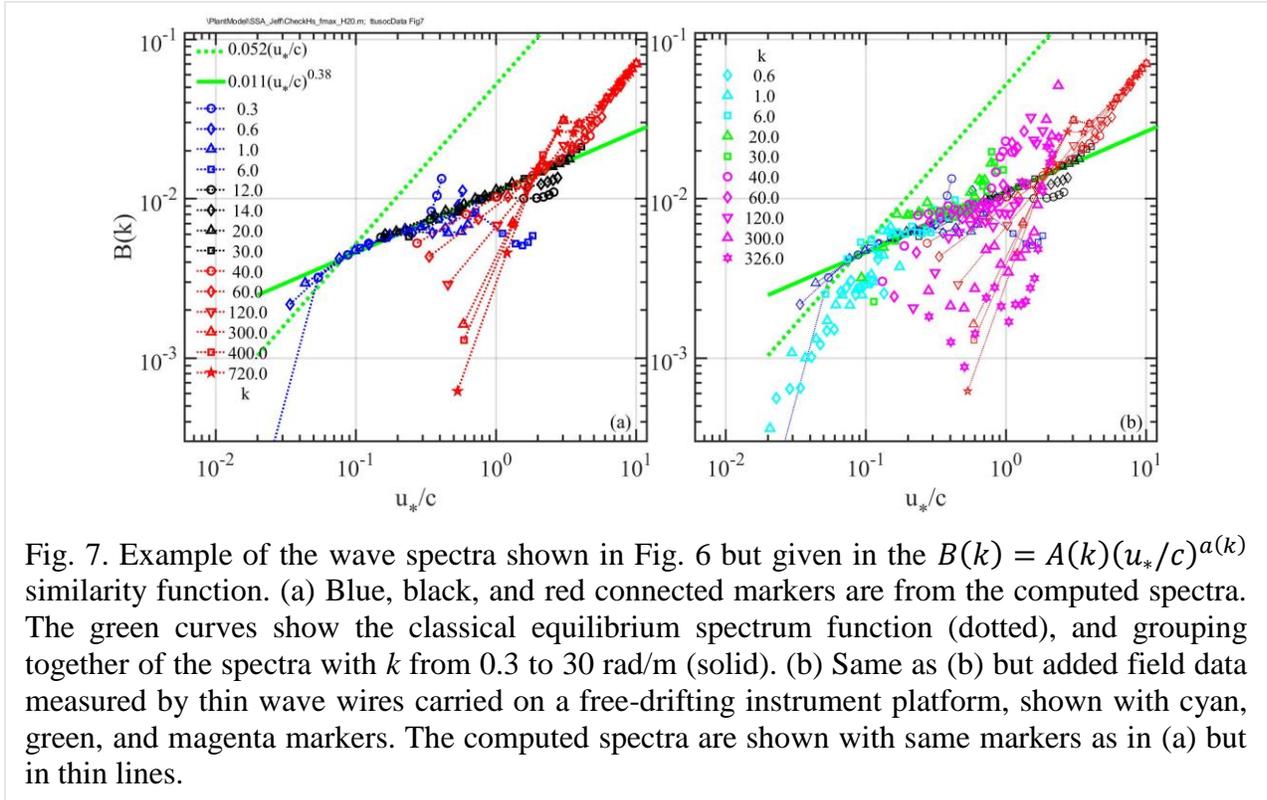

Fig. 7. Example of the wave spectra shown in Fig. 6 but given in the $B(k) = A(k)(u_*/c)^{a(k)}$ similarity function. (a) Blue, black, and red connected markers are from the computed spectra. The green curves show the classical equilibrium spectrum function (dotted), and grouping together of the spectra with $k$ from 0.3 to 30 rad/m (solid). (b) Same as (b) but added field data measured by thin wave wires carried on a free-drifting instrument platform, shown with cyan, green, and magenta markers. The computed spectra are shown with same markers as in (a) but in thin lines.

For convenience, in the following discussion, dmDm is used to represent 0.3 to 30 rad/m $k$ range (wavelengths 0.2 to 20 m). Except for the longest wave component ($k$ = 0.3 rad/m) in the lowest wind ($U_{10}$ = 4 m/s), the displayed dmDm waves are several times shorter than the energy spectral peak wavelength ($k_p$ of each curve is listed inside square brackets of the Figure 6 legend).

Following the argument presented in (Phillips 1984), the source function balance of the dmDm wave action density equation is between the wind input $Q_{in}$ and breaking dissipation $Q_{dis}$ because the nonlinear term is much smaller away from the spectral peak region. The wind input function can be represented by the following formula based on theoretical studies and field data analyses (Plant 1982)

$$Q_{in} = m \left(\frac{u_*}{c}\right)^2 g k^{-3} B(k), \quad m \approx 0.04 \qquad (12)$$

If the dissipation term is expressed as $Q_{dis} \sim f(B)$, and a power function $A_d B^{a_d}$ is assumed for $f(B)$, with $B(k) = A(k)(u_*/c)^{a(k)}$ then $a_d(k) = 1 + 2/a(k)$ and $A_d(k) = mA(k)^{1-a_d(k)}$. Given the observed $a(k) = 0.38$ for dmDm waves, $a_d(k) = 5.26$, that is, the expected $Q_{dis}$ is proportional to $B^{5.26}$, which is much stronger than the cubic dependence as expected from the classical equilibrium spectrum function (Phillips 1985)

$$B_e(k) = 0.052(u_*/c) \qquad (13)$$

which has $a = 1$ and $a_d = 3$; the dotted green curve shows the classical equilibrium spectrum function and it will be further discussed in the last part of this section. For higher wave numbers ($k > \sim 40$ rad/m), the $u_*/c$ dependence is progressively stronger toward the shorter wavelengths (connected red markers) and approaches cubic in mild to moderately high winds (Figure 2 in Hwang 2007, Figure 1a in Hwang 2011). For very high $u_*/c$ region (greater than about 3), the exponents and proportionality coefficients of the $B(k)$ power functions of different wave components converge to same constants as revealed in the analysis of Bragg resonance waves from L-, C-, and Ku-band scatterometer GMFs (Figure 3 in Hwang et al. 2013). In-situ or remote sensing wind and wave related data at high wind speeds are very scattered. Verification and refinement of high-wind wave spectrum properties remain a difficult task. More details on the treatment of the region with $u_*/c > 3$ are described in the supplemental material (H15SpectrumFormulationNote.pdf).

As noted in Hwang (2007), wave breaking is not only an energy sink but also a source of short wave generation, and in fact it is even more effective than wind generation. Growing waves by wind input takes some time but impacts of plunging breaking patches generate a wide band of short waves almost instantaneously. For spilling and surging breakers, the surface deformation is also typically white-noise type with a wide band of wavelengths. The large exponent (5.26) of the "dissipation" power function $f(B)$ discussed in the last paragraph is consistent with the faster response time (higher relaxation rate) of wave breaking as a source term from the relaxation theory view point (Hughes 1978; Plant et al. 1983; Plant and Keller 1983; Hara and Plant 1994; Hwang 1999; Hara et al. 2003). It would be interesting to come up with a physical formulation of breaking as a source of cmDm wave generation and to find out how much such breaking-induced surface deformation contributes to the empirical similarity relation of $B(k) = 0.011(u_*/c)^{0.38}$ observed in the wide wave number band between 0.3 and 30 rad/m (about 0.2 to 20 m wavelengths), and the increase of $u_*/c$ exponent toward the cubic dependence in shorter waves of centimeter lengthscales.

In the field data of free-drifting cmDm wave measurements mentioned above, the resolvable $k$ range is between about 0.6 and 326 rad/m. The field observations are added in Figure 7b together with the computed spectra, which are shown with the same markers as in Figure 7a but in thin lines. These field data are collected in coastal regions in the Northern Gulf of Mexico (various locations within about 40 km from shore). The wind speed range is 2 to 14 m/s and the spectral peak wavelengths are generally much shorter than the spectrum model computations given in Figures 6 and 7: the peak wind-wave periods in the field data are mostly between 3 and 5 s (Figure 1c in Hwang 2008). Despite the limited scope of the geophysical conditions and much more scattered data, the field measurements display some of the prominent features revealed by the synthetic spectral data, including the tight grouping of the wave components between 0.6 and 30 rad/m (cyan and green markers), and the progressively steepening of $u_*/c$ dependence toward the shorter wavelengths (magenta markers). As mentioned earlier, the dotted green line is the classical equilibrium spectrum function (Phillips 1985); the majority of computed and measured data deviates from the equilibrium function.

As a final remark, Figure 8a and 8b compare $B(u_*/c; k)$ and $B(u_*/c_p; k)$ representations of cmDm waves. The parameter $u_*/c_p$ is an alternative expression of the inverse wave age, with the spectral peak wave phase speed normalized by wind friction velocity instead of wind speed. It signifies the effects of spectral peak component on the wave properties. As shown in Figure 5, the effects of spectral peak region diminish quickly toward high frequencies. For investigating the cmDm wave properties, the utility of $u_*/c_p$ as a reference parameter (emphasizing the spectral peak influence) is less informative than $u_*/c$ (emphasizing the spectrally-local balance), this is a particular consideration for formulating parametric wind wave spectrum models extending to capillary-gravity waves.

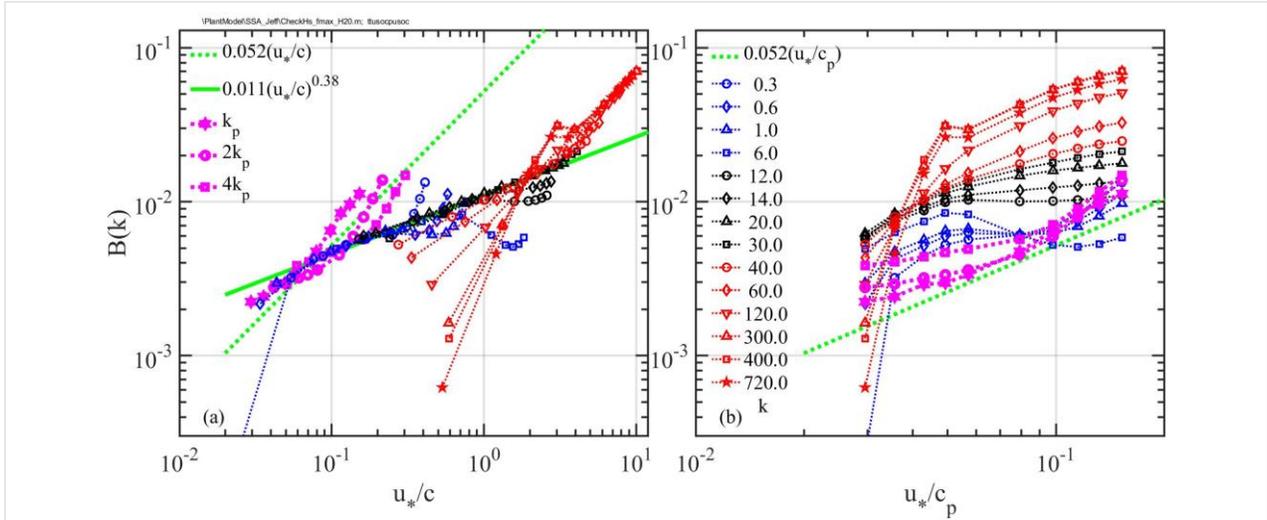

Fig. 8. Comparison of the wave spectra shown in Fig. 6 but given by (a) $B(u_*/c; k)$, and (b) $B(u_*/c_p; k)$.

Also shown in Figure 8 are the results of wave components near the spectral peak: $B(k_p)$, $B(2k_p)$, and $B(4k_p)$, illustrated with magenta markers, and dotted green lines for the equilibrium function (13). In Figure 8a, the added solid green line is the empirical fitting function (11). The equilibrium function is derived under the assumption that the three main sources terms (nonlinear interaction, wind input, and breaking dissipation) are equally important (Phillips 1985). This assumption is more suitable for the wave components near to the spectral peak, and indeed the computed $k_p$, $2k_p$, and $4k_p$ spectral components over a wide wind speed range (4 to 70 m/s) and average wave development stages seem to be close to the equilibrium function. For the wave components much shorter than the spectral peak wavelengths, the source balance is between wind input and breaking dissipation because the nonlinear term is much smaller (Phillips 1984), and $B(u_*/c; k)$ deviates from the equilibrium function established on the balance of three equally-important source terms.

### d. Tilted roughness patch consideration

For Ku and Ka bands, the EM wavelengths range from a few millimeters to a couple centimeters. The Bragg resonance surface waves are capillary and capillary-gravity waves, which have several distinct features when considering microwave scattering applications.

(a) Their generation does not require wind forcing.

(b) They tend to propagate in groups of O(10) wavelets.

(c) In many laboratory experiments, the capillary wave groups are frequently observed to be

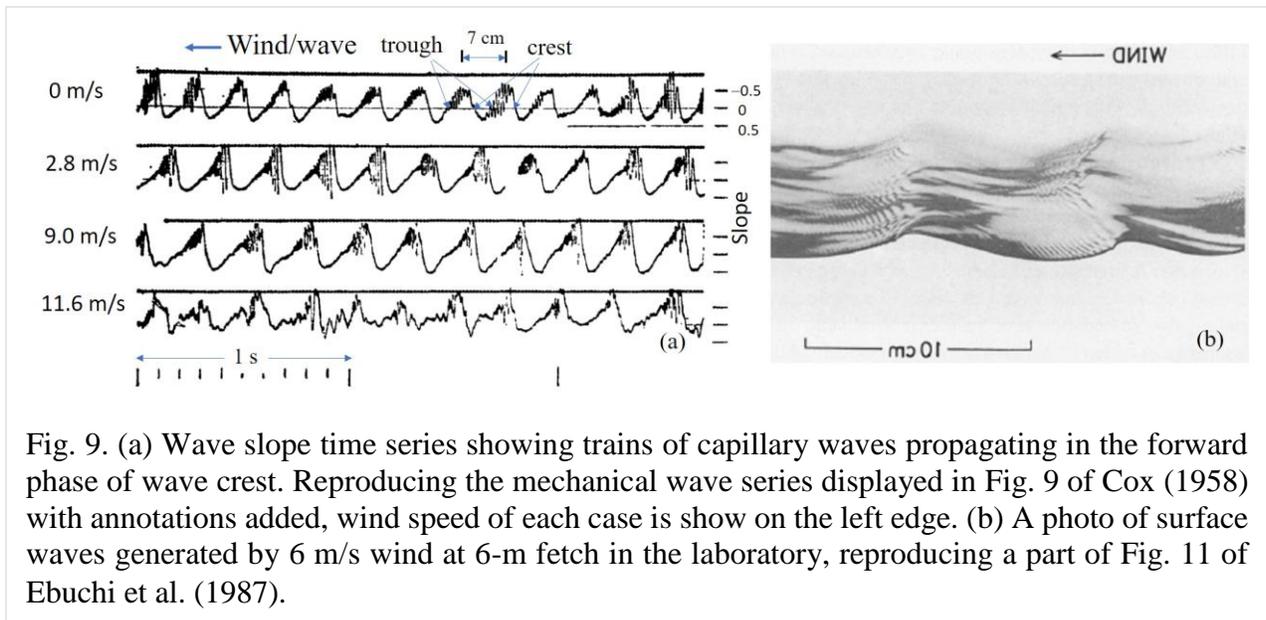

Fig. 9. (a) Wave slope time series showing trains of capillary waves propagating in the forward phase of wave crest. Reproducing the mechanical wave series displayed in Fig. 9 of Cox (1958) with annotations added, wind speed of each case is show on the left edge. (b) A photo of surface waves generated by 6 m/s wind at 6-m fetch in the laboratory, reproducing a part of Fig. 11 of Ebuchi et al. (1987).

in resonant propagation with steep short waves a few centimeters long, and the capillary wave trains are on the forward slope of the steep short wave crest.

Figure 9 shows a few examples from laboratory experiments of capillary wave observations (Cox 1958; Ebuchi et al. 1987). The top trace of Fig. 9a is especially instructive, showing the wave slope time series of capillary wave trains on steep mechanical waves (about 5 Hz frequency and 0.07 m wavelength) without wind forcing. Inspired by this sequence of capillary wave observations, Longuet-Higgins (1963) shows that the primary generation mechanism is localized surface tension force at the sharp crest of the steep mechanical waves. The effect is equivalent to "a travelling disturbance which produces a train of capillary waves ahead of the crest, i.e. on the forward face of the gravity wave." Fig. 9b reproduces a photograph of wind generated short waves in a laboratory wind wave flume (Ebuchi et al. 1987, Fig. 11). The photograph is flipped so the waves are travelling from left to right as those shown in Fig. 9a. There is obvious resemblance between the slope traces in Fig. 9a and the wave image in Fig. 9b photograph. Many more examples of the resonant propagation of groups of capillary waves on steep and short gravity waves a few cm long have be reported (e.g., Chang et al. 1978; Fedorov et al. 1998).

Similar features of discrete groups of capillary waves are frequently observed on the ocean surface as well, two examples are shown in Figure 10. Yellow arrows are added in Fig. 10a to show some such capillary wave groups. The area covered in Fig. 10b is much larger and it is clear to see the inhomogeneous distribution of short waves as reflected in the generally rougher appearance in the crest region and smoother appearance in the trough region, but small areas of rough patches can appear in any phase of the background long waves.

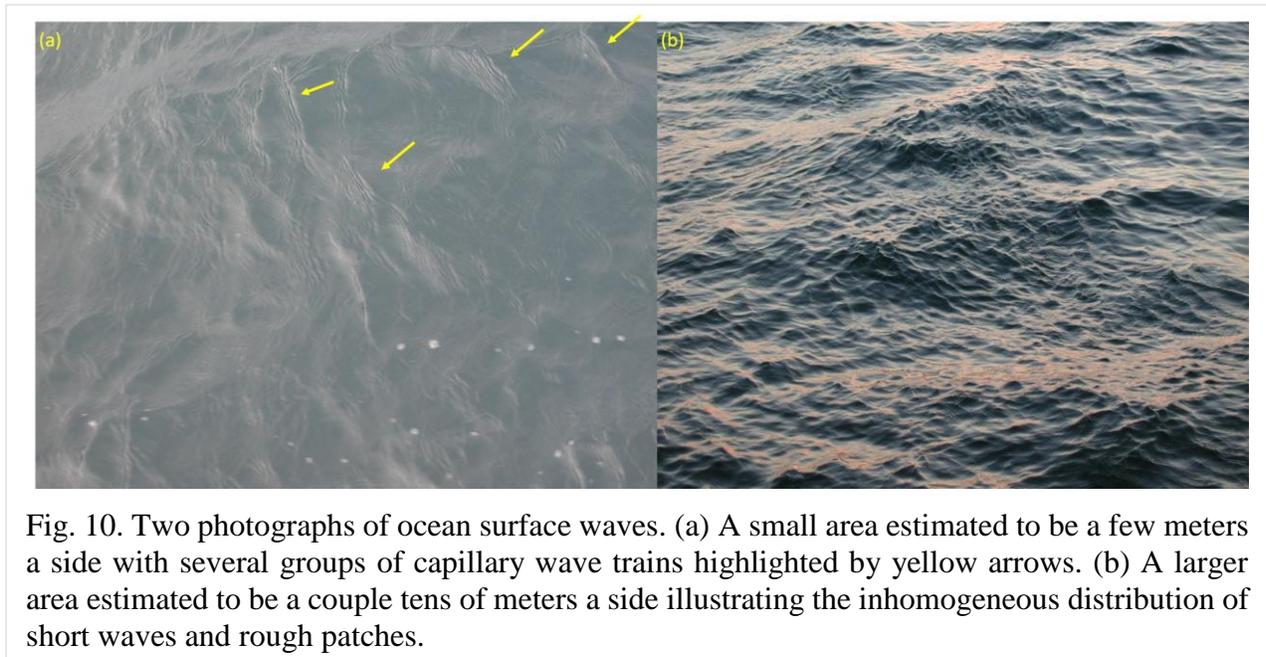

Fig. 10. Two photographs of ocean surface waves. (a) A small area estimated to be a few meters a side with several groups of capillary wave trains highlighted by yellow arrows. (b) A larger area estimated to be a couple tens of meters a side illustrating the inhomogeneous distribution of short waves and rough patches.

Accurate prescription of the nonlinear short gravity waveform is critical to correct computation of the generated capillary wave trains (Longuet-Higgins 1963, 1992, 1995; Fedorov and Melville 1998; Melville and Fedorov 2015; Deike et al. 2015). Implementation of such sophisticated nonlinear computation schemes for microwave scattering computation may not be practical. As shown in Section 2, the tilted specular approach provides a workaround to account for the complicated inhomogeneous distribution and nonlinear nature of ocean surface roughness features important to microwave remote sensing.

## 4. Summary

Mean square slope data presented in (Cox and Munk 1954a,b), with maximum wind speed about 14 m/s, represent the most comprehensive ocean surface roughness measurements for many decades. Microwave ocean remote sensing provides another data source of the ocean surface roughness. Several airborne LPMSS analyses have been reported and the maximum wind extends to about 20 m/s. Here, spaceborne L-, Ku-, and Ka-band NRCS measurements are analyzed to yield LPMSS with 11, 95, and 250 rad/m upper cutoff wave numbers. With the spaceborne platforms, the maximum wind speeds increase considerably, to 80, 29, and 25 m/s, respectively. For the L band, in addition to inversion from the specular NRCS, the LPMSS can be obtained from GPSR Doppler analyses and direct integration using the G18 spectrum function. Results from the three different methods are in good agreement (Figure 4). Presently, the hybrid spectrum function (H18) merges the long-wave function (G18) that is determined by $U_{10}$ and $\omega_\#$, and the short-wave function (H15) that is determined by $u_*/c$. The L band LPMSS has served as the critical data set for determining the average spectral slope in order for the G18 wave spectrum model to yield surface roughness computations in good agreement with the L band LPMSS observations in a wide range of environmental conditions extending to TC winds. Many remote sensing data sets have been used to refine the ocean surface wave spectrum models, especially on the short-wave portion that is important to the determination of the ocean surface roughness. Using the synthetic data derived from spectrum computation over a wide range of wind speeds and average wave development stages (Figures 6 and 7), it is found that over two orders of magnitude of the *k* range (0.3 to 30 rad/m), the wave spectrum can be described by the simple function $B(k) = 0.011(u_*/c)^{0.38}$ for $U_{10} \leq $ ~20 m/s. For shorter wave components, the $u_*/c$ exponent increases toward the cubic dependence for the cm-wavelength components (Figure 7). The analysis presented in this paper suggests that it is important to consider wave breaking as a source of surface roughness generation. The breaking-generated roughness may cover a wideband of wavelengths 20 m and shorter, subject to the condition that the roughness wavelength is several times shorter than the spectral peak wavelength. Finally, under the three-source wave action conservation framework, the classical equilibrium function (Phillips 1985) is established on the balance of three equally-important source terms (nonlinear interaction, wind input, and breaking dissipation) that is more appropriate for describing the wave components near the spectral peak region. For cmDm waves much shorter than the spectral peak wavelengths, the source balance is between wind input and breaking dissipation terms (Phillips 1984), and the spectral preparties deviate from the equilibrium function established on three equally-important source terms (Figure 8).

*Acknowledgements*

This work is supported by the Office of Naval Research under Grant N0001422WX00033, U.S. Naval Research Laboratory Publication JA/7260—22-1030.

*Data Availability Statement.*

All data used in this paper were published in the literature, the specific references were cited in the text. They are also available as a matlab set deposited at https://www.researchgate.net/publication/358280967_MssRemoteSensing.

# Supplements: (a) H15 Spectrum formulation, (b) LUT

## H15 Spectrum formulation

### 1. Low to moderately-high wind conditions

The H15 spectrum function is given by (Hwang and Wang 2004 GRLv31L15301; Hwang 2005 JGRv110C10029)

$$B(k) = A(k)\left(\frac{u_*}{c}\right)^{a(k)} \tag{1}$$

where $B(k) = k^3 S(k)$ is the dimensionless wave spectrum, $S(k)$ is the wave elevation spectrum, $u_*$ is the wind friction velocity, $c$ is the phase speed of the $k$ wave component, $A(k)$ and $a(k)$ are coefficients derived from empirical fitting of field data (Hwang and Wang 2004 GRLv31L15301; Hwang 2005 JGRv110C10029; Hwang 2005 JGRv110C10029; Hwang 2008 JGRv113C12024) and analysis of Bragg resonance waves from L, C, and Ku band GMFs (Hwang et al. 2013 JAOTv30p2168). They are given by analytical functions for three wave number branches: $0 < k \leq k_1$, $k_1 < k \leq k_2$, and $k_2 < k \leq \infty$ rad/m, where $k_1 = 1$ and $k_2 = 500$ rad/m. With greater weighting for the radar data (Ku, C, and L band geophysical model functions) and using the notations $k_{ln}=\ln(k)$, $A_{ln}=\ln(A)$, and $a_{ln}=\ln(a)$, the polynomial fitting for the middle branch ($k_1 \sim k_2$) is (Hwang and Fois 2015 JGRv120p3640):

(a) $A(k)$ and $a(k)$ expressed as fifth order polynomial fitting functions of $k_{ln}=\ln(k)$ for the data range of wave spectrum, $k_1 < k \leq k_2$, ($k$ in rad/m)

$$A_{ln} = -1.6356 \times 10^{-3} k_{ln}^5 + 4.1084 \times 10^{-2} k_{ln}^4 - 3.6789 \times 10^{-1} k_{ln}^3$$
$$+1.3888 k_{ln}^2 - 2.2193 k_{ln} - 3.3179$$

$$a_{ln} = 1.4013 \times 10^{-3} k_{ln}^5 - 2.6997 \times 10^{-2} k_{ln}^4 + 1.5739 \times 10^{-1} k_{ln}^3$$
$$-1.3020 \times 10^{-1} k_{ln}^2 - 7.5202 \times 10^{-1} k_{ln} + 2.3808 \times 10^{-2} \tag{2}$$

(b) $A(k)$ and $a(k)$ approach asymptotically to the equilibrium spectrum with $A_0=5.2\times 10^{-2}$ and $a_0=1$ in the low wave number range, $0 < k \leq k_1$, (Hwang 2008 JGRv113C12024; 2011 JAOTv28p436)

$$A(k) = A_0 \exp\left[\frac{k}{k_1}\left(\ln\frac{A(k_1)}{A_0}\right)\right] = A_0\left(\frac{A(k_1)}{A_0}\right)^{k/k_1}$$

$$a(k) = a_0 \exp\left[\frac{k}{k_1}\left(\ln\frac{a(k_1)}{a_0}\right)\right] = a_0\left(\frac{a(k_1)}{a_0}\right)^{k/k_1}. \tag{3}$$

$A(k_1)$ and $a(k_1)$ are computed with (2) to maintain continuity with the middle branch. The selection of asymptotic values for $A_0$ and $a_0$ are based on extensive research results on the coefficient and exponent of power-law wind speed dependence of the classical equilibrium wind wave spectrum.

(c) $A(k)$ and $a(k)$ approach asymptotically to the high frequency limits of $A_\infty = 2\times10^{-3}$ and $a_\infty = 3$ in the high wave number range, $k_2 < k \leq \infty$, (Hwang 2008 JGRv113C12024; 2011 JAOTv28p436)

$$A(k) = A_\infty exp\left[\frac{k_2}{k}\left(ln\frac{A(k_2)}{A_\infty}\right)\right] = A_\infty \left(\frac{A(k_2)}{A_\infty}\right)^{k_2/k}$$

$$a(k) = a_\infty exp\left[\frac{k_2}{k}\left(ln\frac{a(k_2)}{a_\infty}\right)\right] = a_\infty \left(\frac{a(k_2)}{a_\infty}\right)^{k_2/k}. \quad (4)$$

$A(k_2)$ and $a(k_2)$ are computed with (2) to maintain continuity with the middle branch. The choice of $a_\infty$ value is based on the numerous measurements of the wind speed dependence of microwave radar cross sections and radiometer brightness temperatures at different incidence angles and various microwave frequencies. The choice of $A_\infty$ is more difficult due to a lack of reference information and $2\times10^{-3}$ is used to complete the parameterization.

The (inverse) wave age parameter $\omega_\# = \omega_p U_{10}/g$, which is $U_{10}/c_p$ for the deep-water wave condition, can be introduced in the low wavenumber branch by multiplying the computed $B(k)$ with a Gaussian shape function designed in a similar fashion of the Pierson-Moskowitz spectrum (Pierson and Moskowitz 1964 JGRv69p5181), as described in Hwang et al. (2013 JAOTv30p2168):

$$F_G = \begin{cases} exp\left[-\left(\frac{k_p}{k}\right)^2 - 1\right], & k < k_p \\ 1, & k \geq k_p \end{cases}. \quad (5)$$

The peak enhancement factor that is an important feature of nonlinear wave-wave interaction in the JONSWAP spectrum (Hasselmann et al. 1973 DHZvA8p1) can also be incorporated:

$$F_p(k) = \gamma^{exp\left[\frac{-(k-k_p)}{2\sigma^2 k_p}\right]}, \sigma = \begin{cases} \sigma_1, k < k_p \\ \sigma_2, k \geq k_p \end{cases}. \quad (6)$$

From field experiments, the mean values of $\gamma$ and $\sigma$ are about 3 and 0.1; the difference between $\sigma_1$ and $\sigma_2$ is small (Fig. 2.8 in Hasselmann et al. 1973 DHZvA8p1). Although these peak modifications are important feature of the wind wave energy spectrum, their influence on microwave backscattering and brightness temperature computations is negligible.

## 2. High wind condition

The roughness spectra in high winds ($u_*/c$ greater than about 3) derived from the three (L, C, and Ku) scatterometer GMFs tend to converge to $(u_*/c)^{0.75}$ similarity form (Figure 3 in Hwang et al. 2013 JAOTv30p2168). Because the minimum surface wave phase speed $c_{min} = 0.23$ m/s for the $k_{min} = 369$ rad/m component, $u_*/c$ greater than about 3 corresponding to $u_*$ greater than about 0.69, or $U_{10}$ greater than about 16 m/s. For the nominal L, C, Ku, and Ka wave numbers 33.5, 126, 293, and 754 rad/m, the corresponding phase speeds are respectively 0.54, 0.30, 0.23, and 0.26 m/s, $u_* = 3c$ are 1.63, 0.89, 0.70, and 0.78 m/s, and the corresponding wind speeds are 35, 30, 16, and 18 m/s.

To extend the roughness spectrum to high wind speeds, we may use this observed similarity asymptote behavior. Specifically, the following modification can be implemented for the $k$ components satisfying $u_*/c \geq 3$,

$$B_h\left(\frac{u_*}{c}; k\right) = A_h(k) \left(\frac{u_*}{c}\right)^{a_h(k)} \tag{7a}$$

$$a_h(k) = 0.75 \tag{7b}$$

$$A_h(k) = A_{11}(k_m)(3)^{a_{11}(k_m)-0.75} \tag{7c}$$

where subscript $h$ indicates the quantities for $u_*/c \geq 3$, $A_{11}(k_m)$ and $a_{11}(k_m)$ are calculated with the algorithm described in Section 1 and $k_m$ represents the matching wavenumber at $(u_*/c)_m = 3$, there are 0, 1 or 2 values of $k_m$ because the wave phase speed $c$ has a minimum value of $c_{min} \approx 0.23$ m/s at $k_{min} \approx 369$ rad/m; the modification occurs in the neighborhood surrounding $k_{min}$. In the computer code, only when the number of spectral components with ($u_*/c$ greater than about 3) exceeds 3, the Eq. 7 modification takes place.

## 3. Post-2015 modifications

In later implementations of the surface roughness spectrum used for microwave remote sensing computations, the low frequency portion of the H15 is replaced by the G18 spectrum function, and the hybrid spectrum is called the H18. There is a small modification of the original H15 component adopted in the H18 spectrum model. The modification is prompted by the much lower spectral levels in the low wavenumber portion of the G18 spectrum compared to the original H15 low wavenumber portion in high winds, thus it requires a reassessment of the effects on the vertical transmit vertical receive (VV) NRCS computations. Using a similar modest goal of achieving within 2 dB difference between the modeled NRCSs and the L, C and Ku band GMFs, the only change in the original H15 component used in H18 is the exponent in the high wind switch described in Eq. (7) of Hwang et al. (2013 JAOTv30p2168), which is also Eq. (7) in this supplement: the exponent is modified from the constant value 0.75 [in Eq. (7b) and (7c)] to

$$a_h(k) = 0.6 + 0.6\tanh(0.02k) \tag{8b}$$

More details are given in Hwang (2019 TGRSv57p2766).

To further smooth the transition, the ($A_h$, $a_h$) on both ends of the spectral components with ($u_*/c$ greater than about 3) can be modified by making use of the values of the ($A$, $a$) values just outside of the spectral components with ($u_*/c$ greater than about 3). Denoting ($A_{left}$, $a_{left}$) and ($A_{rightt}$, $a_{right}$) as the left- and right-outside values of ($A$, $a$), the most-current code uses the following formulas:

$$a_h(k) = a_{left} + (1.2 - a_{left})\tanh(0.02k) \tag{9b}$$

$$A_h(k) = A_{left}(u_*/c)^{(a_{left} - a_h(k))} \tag{9c}$$

Here *left* and *right* refer to the positions on the wave number axis of the spectral components with ($u_*/c$ greater than about 3). To remove the kink at the high wave number tail segment, i.e., to the right-side of ($u_*/c$ greater than about 3), the $A$ values of the tail segment ($A_{tail}$) is multiplied with a factor

$$F_{tail} = \left[A_{left}(u_*/c)^{a_{left}}\right] / \left[A_{right}(u_*/c)^{a_{right}}\right] \tag{10}$$

The H15 short wave portion of the wave spectrum used in the microwave computations presented in the paper are based on this most-current implementation.

In-situ or remote sensing wind and wave related data at high wind speeds are very scattered. Verification and refinement of high-wind wave spectrum properties remain a difficult task.

## 4. Finite depth effects and swell components

Wave shoaling effects can be introduced in the G18 spectrum ($G_{fd}$/$H18_{fd}$) as discussed in Hwang (2022 JPO in press, doi: 10.1175/JPO-D-21-0258.1 published online). Swell components can also be incorporated (G20/H20, with several swell spectrum options offered) as discussed in Hwang et al. (2022 TGRSv60 Art. no. 4204512, doi: 10.1109/10.1109/TGRS.2021.3118995).

| U10 m/s | sigL dB | 100s2_L G18 | sigL dB | 100s2_L H18 | sigC dB | 100s2_C | sigX dB | 100s2_X | sigKu dB | 100s2_Ku | sigKa dB | 100s2_Ka |
|---|---|---|---|---|---|---|---|---|---|---|---|---|
| 1 | 25.12 | 0.17 | 22.34 | 0.32 | 18.47 | 0.73 | 17.79 | 0.84 | 17.56 | 0.87 | 16.89 | 0.90 |
| 2 | 19.64 | 0.59 | 19.22 | 0.65 | 16.36 | 1.18 | 15.71 | 1.35 | 15.45 | 1.41 | 14.68 | 1.50 |
| 3 | 17.28 | 1.01 | 17.42 | 0.98 | 14.94 | 1.63 | 14.30 | 1.86 | 14.04 | 1.95 | 13.22 | 2.09 |
| 4 | 16.26 | 1.28 | 16.27 | 1.28 | 14.04 | 2.01 | 13.40 | 2.29 | 13.12 | 2.40 | 12.22 | 2.63 |
| 5 | 15.44 | 1.55 | 15.37 | 1.57 | 13.28 | 2.39 | 12.65 | 2.71 | 12.36 | 2.86 | 11.41 | 3.17 |
| 6 | 14.91 | 1.75 | 14.79 | 1.80 | 12.78 | 2.68 | 12.15 | 3.05 | 11.84 | 3.23 | 10.81 | 3.64 |
| 7 | 14.44 | 1.95 | 14.28 | 2.02 | 12.34 | 2.97 | 11.69 | 3.39 | 11.37 | 3.59 | 10.28 | 4.11 |
| 8 | 14.09 | 2.11 | 13.90 | 2.20 | 11.99 | 3.21 | 11.34 | 3.67 | 11.00 | 3.91 | 9.84 | 4.55 |
| 9 | 13.76 | 2.27 | 13.56 | 2.38 | 11.68 | 3.46 | 11.01 | 3.96 | 10.66 | 4.22 | 9.43 | 4.99 |
| 10 | 13.50 | 2.41 | 13.28 | 2.54 | 11.42 | 3.67 | 10.74 | 4.21 | 10.37 | 4.51 | 9.07 | 5.42 |
| 11 | 13.26 | 2.55 | 13.01 | 2.70 | 11.17 | 3.88 | 10.48 | 4.46 | 10.11 | 4.79 | 8.74 | 5.85 |
| 12 | 13.05 | 2.67 | 12.79 | 2.84 | 10.96 | 4.06 | 10.26 | 4.69 | 9.87 | 5.05 | 8.43 | 6.27 |
| 13 | 12.81 | 2.82 | 12.58 | 2.98 | 10.76 | 4.26 | 10.04 | 4.92 | 9.65 | 5.31 | 8.14 | 6.69 |
| 14 | 12.57 | 2.99 | 12.38 | 3.12 | 10.57 | 4.44 | 9.85 | 5.14 | 9.44 | 5.56 | 7.86 | 7.12 |
| 15 | 12.33 | 3.15 | 12.18 | 3.26 | 10.39 | 4.63 | 9.65 | 5.37 | 9.23 | 5.82 | 7.60 | 7.56 |
| 16 | 12.11 | 3.31 | 12.01 | 3.39 | 10.22 | 4.80 | 9.48 | 5.58 | 9.05 | 6.06 | 7.35 | 7.98 |
| 17 | 11.91 | 3.47 | 11.84 | 3.53 | 10.06 | 4.98 | 9.31 | 5.79 | 8.87 | 6.31 | 7.11 | 8.41 |
| 18 | 11.82 | 3.54 | 11.72 | 3.62 | 9.94 | 5.11 | 9.17 | 5.96 | 8.72 | 6.51 | 6.96 | 8.67 |
| 19 | 11.72 | 3.61 | 11.59 | 3.72 | 9.81 | 5.25 | 9.03 | 6.14 | 8.57 | 6.71 | 6.81 | 8.95 |
| 20 | 11.73 | 3.61 | 11.52 | 3.78 | 9.72 | 5.35 | 8.92 | 6.27 | 8.45 | 6.87 | 6.68 | 9.19 |
| 21 | 11.72 | 3.61 | 11.43 | 3.86 | 9.63 | 5.46 | 8.81 | 6.41 | 8.33 | 7.05 | 6.54 | 9.45 |
| 22 | 11.70 | 3.62 | 11.35 | 3.92 | 9.54 | 5.56 | 8.71 | 6.54 | 8.21 | 7.21 | 6.40 | 9.69 |
| 23 | 11.69 | 3.63 | 11.27 | 3.99 | 9.45 | 5.66 | 8.60 | 6.68 | 8.09 | 7.37 | 6.27 | 9.93 |
| 24 | 11.66 | 3.64 | 11.19 | 4.06 | 9.36 | 5.76 | 8.49 | 6.81 | 7.98 | 7.52 | 6.13 | 10.18 |
| 25 | 11.63 | 3.66 | 11.11 | 4.13 | 9.27 | 5.86 | 8.39 | 6.94 | 7.86 | 7.68 | 6.00 | 10.44 |
| 26 | 11.60 | 3.68 | 11.04 | 4.19 | 9.18 | 5.95 | 8.28 | 7.07 | 7.75 | 7.84 | 5.85 | 10.72 |
| 27 | 11.56 | 3.71 | 10.96 | 4.27 | 9.10 | 6.05 | 8.18 | 7.20 | 7.63 | 8.00 | 5.70 | 11.00 |
| 28 | 11.53 | 3.74 | 10.88 | 4.33 | 9.01 | 6.15 | 8.07 | 7.33 | 7.52 | 8.16 | 5.57 | 11.25 |
| 29 | 11.49 | 3.76 | 10.81 | 4.40 | 8.93 | 6.25 | 7.97 | 7.45 | 7.40 | 8.31 | 5.43 | 11.51 |
| 30 | 11.44 | 3.79 | 10.73 | 4.47 | 8.84 | 6.34 | 7.86 | 7.58 | 7.29 | 8.46 | 5.29 | 11.76 |
| 31 | 11.39 | 3.83 | 10.65 | 4.54 | 8.76 | 6.43 | 7.76 | 7.71 | 7.17 | 8.62 | 5.15 | 12.02 |
| 32 | 11.35 | 3.86 | 10.58 | 4.61 | 8.68 | 6.52 | 7.65 | 7.83 | 7.06 | 8.76 | 5.01 | 12.27 |
| 33 | 11.30 | 3.90 | 10.51 | 4.68 | 8.60 | 6.62 | 7.54 | 7.95 | 6.94 | 8.91 | 4.87 | 12.52 |
| 34 | 11.24 | 3.94 | 10.44 | 4.74 | 8.51 | 6.71 | 7.44 | 8.07 | 6.83 | 9.06 | 4.73 | 12.77 |
| 35 | 11.19 | 3.98 | 10.37 | 4.81 | 8.43 | 6.80 | 7.33 | 8.20 | 6.71 | 9.21 | 4.59 | 13.02 |
| 36 | 11.14 | 4.02 | 10.31 | 4.87 | 8.37 | 6.87 | 7.25 | 8.30 | 6.63 | 9.33 | 4.49 | 13.23 |
| 37 | 11.09 | 4.06 | 10.25 | 4.94 | 8.31 | 6.95 | 7.18 | 8.40 | 6.54 | 9.46 | 4.38 | 13.44 |
| 38 | 11.04 | 4.10 | 10.19 | 5.00 | 8.25 | 7.02 | 7.11 | 8.48 | 6.47 | 9.54 | 4.31 | 13.55 |
| 39 | 10.99 | 4.15 | 10.13 | 5.06 | 8.20 | 7.09 | 7.04 | 8.56 | 6.40 | 9.63 | 4.24 | 13.66 |
| 40 | 10.93 | 4.20 | 10.07 | 5.12 | 8.13 | 7.17 | 6.96 | 8.66 | 6.32 | 9.75 | 4.14 | 13.88 |
| 41 | 10.88 | 4.24 | 10.01 | 5.18 | 8.07 | 7.25 | 6.88 | 8.77 | 6.23 | 9.88 | 4.03 | 14.10 |
| 42 | 10.83 | 4.29 | 9.95 | 5.25 | 8.01 | 7.32 | 6.82 | 8.85 | 6.16 | 9.97 | 3.96 | 14.20 |
| 43 | 10.77 | 4.34 | 9.89 | 5.31 | 7.96 | 7.39 | 6.75 | 8.93 | 6.09 | 10.06 | 3.88 | 14.31 |
| 44 | 10.72 | 4.39 | 9.84 | 5.38 | 7.90 | 7.46 | 6.68 | 9.01 | 6.02 | 10.14 | 3.81 | 14.42 |
| 45 | 10.66 | 4.44 | 9.78 | 5.44 | 7.84 | 7.54 | 6.61 | 9.09 | 5.94 | 10.23 | 3.74 | 14.53 |

| | | | | | | | | | | | |
|---|---|---|---|---|---|---|---|---|---|---|---|
| 46 | 10.60 | 4.49 | 9.72 | 5.51 | 7.78 | 7.62 | 6.52 | 9.20 | 5.85 | 10.36 | 3.63 | 14.75 |
| 47 | 10.55 | 4.54 | 9.66 | 5.57 | 7.72 | 7.70 | 6.44 | 9.30 | 5.76 | 10.49 | 3.51 | 14.97 |
| 48 | 10.49 | 4.60 | 9.61 | 5.64 | 7.66 | 7.77 | 6.37 | 9.38 | 5.69 | 10.57 | 3.44 | 15.08 |
| 49 | 10.43 | 4.65 | 9.55 | 5.71 | 7.60 | 7.85 | 6.30 | 9.47 | 5.61 | 10.66 | 3.36 | 15.18 |
| 50 | 10.38 | 4.71 | 9.49 | 5.77 | 7.54 | 7.93 | 6.21 | 9.57 | 5.52 | 10.79 | 3.24 | 15.41 |
| 51 | 10.32 | 4.76 | 9.44 | 5.84 | 7.47 | 8.01 | 6.12 | 9.68 | 5.42 | 10.92 | 3.13 | 15.65 |
| 52 | 10.26 | 4.82 | 9.38 | 5.91 | 7.41 | 8.09 | 6.05 | 9.76 | 5.34 | 11.01 | 3.04 | 15.75 |
| 53 | 10.20 | 4.88 | 9.32 | 5.98 | 7.36 | 8.17 | 5.97 | 9.85 | 5.26 | 11.10 | 2.96 | 15.85 |
| 54 | 10.14 | 4.93 | 9.27 | 6.04 | 7.30 | 8.24 | 5.90 | 9.93 | 5.18 | 11.18 | 2.87 | 15.96 |
| 55 | 10.09 | 4.99 | 9.21 | 6.11 | 7.24 | 8.32 | 5.82 | 10.01 | 5.10 | 11.27 | 2.79 | 16.06 |
| 56 | 10.03 | 5.05 | 9.16 | 6.18 | 7.17 | 8.40 | 5.73 | 10.12 | 5.00 | 11.41 | 2.66 | 16.30 |
| 57 | 9.97 | 5.11 | 9.10 | 6.25 | 7.11 | 8.49 | 5.64 | 10.24 | 4.90 | 11.55 | 2.54 | 16.54 |
| 58 | 9.91 | 5.17 | 9.04 | 6.32 | 7.05 | 8.57 | 5.55 | 10.32 | 4.81 | 11.63 | 2.44 | 16.64 |
| 59 | 9.85 | 5.23 | 8.99 | 6.39 | 6.99 | 8.64 | 5.47 | 10.40 | 4.72 | 11.72 | 2.35 | 16.75 |
| 60 | 9.79 | 5.30 | 8.93 | 6.46 | 6.93 | 8.72 | 5.39 | 10.49 | 4.63 | 11.81 | 2.25 | 16.85 |
| 61 | 9.74 | 5.36 | 8.88 | 6.54 | 6.87 | 8.80 | 5.30 | 10.57 | 4.54 | 11.90 | 2.15 | 16.95 |
| 62 | 9.68 | 5.42 | 8.82 | 6.61 | 6.81 | 8.88 | 5.21 | 10.65 | 4.44 | 11.98 | 2.05 | 17.06 |
| 63 | 9.62 | 5.49 | 8.77 | 6.68 | 6.75 | 8.96 | 5.12 | 10.74 | 4.35 | 12.07 | 1.94 | 17.16 |
| 64 | 9.56 | 5.55 | 8.71 | 6.76 | 6.68 | 9.05 | 5.02 | 10.85 | 4.23 | 12.21 | 1.80 | 17.40 |
| 65 | 9.50 | 5.62 | 8.66 | 6.83 | 6.62 | 9.14 | 4.92 | 10.97 | 4.11 | 12.35 | 1.66 | 17.65 |
| 66 | 9.44 | 5.68 | 8.61 | 6.90 | 6.56 | 9.22 | 4.82 | 11.05 | 4.00 | 12.44 | 1.54 | 17.75 |
| 67 | 9.39 | 5.75 | 8.55 | 6.98 | 6.49 | 9.30 | 4.72 | 11.14 | 3.90 | 12.53 | 1.43 | 17.86 |
| 68 | 9.33 | 5.82 | 8.50 | 7.05 | 6.43 | 9.38 | 4.62 | 11.23 | 3.79 | 12.62 | 1.31 | 17.96 |
| 69 | 9.27 | 5.89 | 8.44 | 7.13 | 6.37 | 9.46 | 4.52 | 11.31 | 3.67 | 12.71 | 1.18 | 18.07 |
| 70 | 9.21 | 5.96 | 8.39 | 7.21 | 6.30 | 9.54 | 4.42 | 11.40 | 3.56 | 12.80 | 1.05 | 18.17 |
| 71 | 9.15 | 6.03 | 8.33 | 7.28 | 6.24 | 9.62 | 4.31 | 11.49 | 3.43 | 12.89 | 0.92 | 18.27 |
| 72 | 9.09 | 6.09 | 8.28 | 7.36 | 6.17 | 9.71 | 4.19 | 11.61 | 3.29 | 13.04 | 0.75 | 18.53 |
| 73 | 9.04 | 6.17 | 8.23 | 7.44 | 6.10 | 9.81 | 4.06 | 11.73 | 3.14 | 13.18 | 0.57 | 18.78 |
| 74 | 8.98 | 6.24 | 8.17 | 7.52 | 6.04 | 9.89 | 3.95 | 11.81 | 3.01 | 13.28 | 0.42 | 18.89 |
| 75 | 8.92 | 6.31 | 8.12 | 7.59 | 5.97 | 9.97 | 3.82 | 11.90 | 2.87 | 13.37 | 0.27 | 18.99 |
| 76 | 8.86 | 6.38 | 8.07 | 7.67 | 5.91 | 10.06 | 3.70 | 11.99 | 2.73 | 13.46 | 0.10 | 19.10 |
| 77 | 8.80 | 6.45 | 8.01 | 7.75 | 5.84 | 10.14 | 3.57 | 12.08 | 2.57 | 13.55 | -0.06 | 19.20 |
| 78 | 8.75 | 6.53 | 7.96 | 7.83 | 5.77 | 10.23 | 3.44 | 12.17 | 2.42 | 13.65 | -0.24 | 19.31 |
| 79 | 8.69 | 6.60 | 7.91 | 7.91 | 5.70 | 10.32 | 3.30 | 12.26 | 2.25 | 13.74 | -0.42 | 19.41 |
| 80 | 8.63 | 6.68 | 7.85 | 8.00 | 5.63 | 10.41 | 3.15 | 12.38 | 2.06 | 13.89 | -0.65 | 19.68 |
| 81 | 8.57 | 6.75 | 7.80 | 8.08 | 5.56 | 10.51 | 2.98 | 12.51 | 1.87 | 14.04 | -0.88 | 19.94 |
| 82 | 8.51 | 6.83 | 7.75 | 8.16 | 5.49 | 10.59 | 2.83 | 12.60 | 1.68 | 14.14 | -1.09 | 20.05 |
| 83 | 8.46 | 6.91 | 7.70 | 8.24 | 5.42 | 10.68 | 2.67 | 12.69 | 1.48 | 14.23 | -1.31 | 20.16 |
| 84 | 8.40 | 6.98 | 7.64 | 8.32 | 5.34 | 10.77 | 2.50 | 12.78 | 1.27 | 14.33 | -1.55 | 20.26 |
| 85 | 8.34 | 7.06 | 7.59 | 8.41 | 5.27 | 10.86 | 2.32 | 12.88 | 1.05 | 14.42 | -1.79 | 20.37 |
| 86 | 8.28 | 7.14 | 7.54 | 8.49 | 5.20 | 10.95 | 2.14 | 12.97 | 0.82 | 14.52 | -2.06 | 20.48 |
| 87 | 8.23 | 7.22 | 7.49 | 8.58 | 5.12 | 11.04 | 1.94 | 13.06 | 0.57 | 14.61 | -2.33 | 20.59 |
| 88 | 8.17 | 7.30 | 7.43 | 8.66 | 5.05 | 11.13 | 1.74 | 13.16 | 0.30 | 14.71 | -2.63 | 20.69 |
| 89 | 8.11 | 7.38 | 7.38 | 8.75 | 4.97 | 11.22 | 1.52 | 13.25 | 0.02 | 14.81 | -2.95 | 20.80 |
| 90 | 8.05 | 7.46 | 7.33 | 8.84 | 4.89 | 11.32 | 1.28 | 13.38 | -0.31 | 14.97 | -3.32 | 21.08 |
| 91 | 8.00 | 7.54 | 7.28 | 8.92 | 4.81 | 11.42 | 1.03 | 13.51 | -0.65 | 15.12 | -3.72 | 21.36 |
| 92 | 7.94 | 7.63 | 7.22 | 9.01 | 4.73 | 11.51 | 0.77 | 13.61 | -1.01 | 15.22 | -4.12 | 21.47 |

| 93 | 7.88 | 7.71 | 7.17 | 9.10 | 4.65 | 11.60 | 0.50 | 13.70 | -1.40 | 15.32 | -4.54 | 21.58 |
| 94 | 7.82 | 7.79 | 7.12 | 9.19 | 4.57 | 11.69 | 0.20 | 13.80 | -1.82 | 15.42 | -5.01 | 21.69 |
| 95 | 7.77 | 7.88 | 7.07 | 9.28 | 4.48 | 11.79 | -0.12 | 13.90 | -2.28 | 15.52 | -5.52 | 21.80 |
| 96 | 7.71 | 7.96 | 7.01 | 9.37 | 4.40 | 11.88 | -0.46 | 13.99 | -2.80 | 15.62 | -6.08 | 21.91 |
| 97 | 7.65 | 8.05 | 6.96 | 9.46 | 4.31 | 11.98 | -0.83 | 14.09 | -3.37 | 15.72 | -6.71 | 22.02 |
| 98 | 7.60 | 8.13 | 6.91 | 9.55 | 4.22 | 12.07 | -1.23 | 14.19 | -4.02 | 15.82 | -7.41 | 22.13 |
| 99 | 7.54 | 8.22 | 6.86 | 9.64 | 4.13 | 12.17 | -1.67 | 14.29 | -4.75 | 15.93 | -8.20 | 22.25 |